# AI in 6G: Energy-Efficient Distributed Machine Learning for Multilayer Heterogeneous Networks

Mohammad Arif Hossain, Abdullah Ridwan Hossain, and Nirwan Ansari

*Department of Electrical and Computer Engineering, New Jersey Institute of Technology*

*Newark, NJ, 07102 USA*

*mh624@njit.edu; arh24@njit.edu; nirwan.ansari@njit.edu*

*Abstract*— **Adept network management is key for supporting extremely heterogeneous applications with stringent quality of service (QoS) requirements; this is more so when envisioning the complex and ultra-dense 6G mobile heterogeneous network (HetNet). From both the environmental and economical perspectives, non-homogeneous QoS demands obstruct the minimization of the energy footprints and operational costs of the envisioned robust networks. As such, network intelligentization is expected to play an essential role in the realization of such sophisticated aims. The fusion of artificial intelligence (AI) and mobile networks will allow for the dynamic and automatic configuration of network functionalities. Machine learning (ML), one of the backbones of AI, will be instrumental in forecasting changes in network loads and resource utilization, estimating channel conditions, optimizing network slicing, and enhancing security and encryption. However, it is well known that ML tasks themselves incur massive computational burdens and energy costs. To overcome such obstacles, we propose a novel layer-based HetNet architecture which optimally distributes tasks associated with different ML approaches across network layers and entities; such a HetNet boasts multiple access schemes as well as device-to-device (D2D) communications to enhance energy efficiency via collaborative learning and communications.**

# I. INTRODUCTION

The envisioned future applications and services have introduced massive challenges for 6G networks. Numerous research endeavors have improved existing techniques while proposing novel techniques to facilitate support for stringent and diverse Quality of Service (QoS) requirements. Nevertheless, it has been made clear that even 5G networks will eventually strain under the more advanced envisioned use cases [1]. 6G networks are expected to support a connectivity density of $10^7$ devices\per\km\squared} which is ten times the connectivity density of 5G. The peak throughput is expected to be at least 1 Tb/s whereas in 5G, it is 20 Gb/s. Furthermore, the spectral and energy efficiencies are expected to be 5-10 and 10-100 times higher than those of 5G, respectively. The latency and user equipment (UE) mobility are envisioned to be 10-100 μs and at least 1000 km/h, correspondingly [2], [3]. Hence, the formal 5G use cases referred to as enhanced mobile broadband (eMBB), ultra-reliable and low latency communications (uRLLC), and massive machine type communications (mMTC) are repackaged as further enhanced mobile broadband (feMBB), extremely reliable and low latency communications (ERLLC), and ultra-massive machine type communications (umMTC), correspondingly. 6G literature has even further proposed additional use cases: extremely low-power communications (ELPC), long-distance and high-mobility communications (LDHMC), massive uRLLC (muRLLC), and mobile broadband reliable and low latency communications (MBRLLC) [4]. While their QoS parameters have yet to be solidified, the labelling alone points towards the increasing tightening of QoS requirements.

Such use cases will force major changes in network architecture and management; one such fundamental change will be the fusion of artificial intelligence (AI) [5]-[7] that will likely supersede earlier attempts [8]. On the contrary, it is not simply the supplementing of a network with AI but rather making AI a core basis of a network because AI will be instrumental in enhancing different network functions for 6G such as resource utilization, estimating channel conditions, optimizing network slicing, enhancing encryption and security, and more [9]-[11]. Specifically, it is machine learning (ML) that will enhance the above to provision various 6G services in an efficient manner. However, ML tasks require massive computational power which translates to significant energy consumption of not only the network but also the UEs as well. Consequently, the decentralization of ML in a network will dampen the power consumption and computational load of the network.

In this work, we envision a MUltilayer diStributed archItecture for energy efficient machine learning enabled Communications (**MUSIC**) for 6G to carry out distributed collaborative learning across several layers of the network. To further minimize the communication costs of the network, we also propose the use of several multiple access (MA) schemes at the device layer such as orthogonal multiple access (OMA) and non-orthogonal multiple access (NOMA). Moreover, we employ device-to-device (D2D) communications to facilitate the distribution of ML tasks among collaborating UEs at the device layer. Such an ML architecture is envisioned to be robust such that it facilitates the intellectualization of the 6G network with respect to: network management, resource allocation, channel estimation, network slicing orchestration, etc. Not limited to the above, it should likewise provide an ML execution environment for other tasks which can benefit from ML. We elaborate on such a novel architecture in the following sections.

## II. Multi-layer Distributed ML-Enabled 6G Heterogeneous Network Architecture

### A. 6G Heterogeneous Network

We present our proposed heterogeneous network (HetNet) architecture supporting several MA schemes in Fig. 1. According to our architecture, the macro base-station (MBS) can utilize sub-6 GHz, mmWave, and THz frequency bands for communication. As the network transverses radially outward from the MBS, the micro base-station ($\mu$BS) and access points (APs) provide small cell connectivity by utilizing mmWave and THz bands only. Each cell supports both OMA and NOMA to enhance the spectral and energy efficiencies of the system (detailed in Section IV). Furthermore, both OMA and NOMA can utilize flexible subcarrier spacing to enhance throughput, minimize communication power, and shorten latency. UEs can also form D2D networks among themselves and a UE may be designated as a relay between two UEs, between the BS and another UE, or between an AP and another UE. Next, we strategically integrate computational servers at various layers of the HetNet to endow it with ML capabilities and optimize the efficiency of the HetNet from the perspective of both communications and ML.

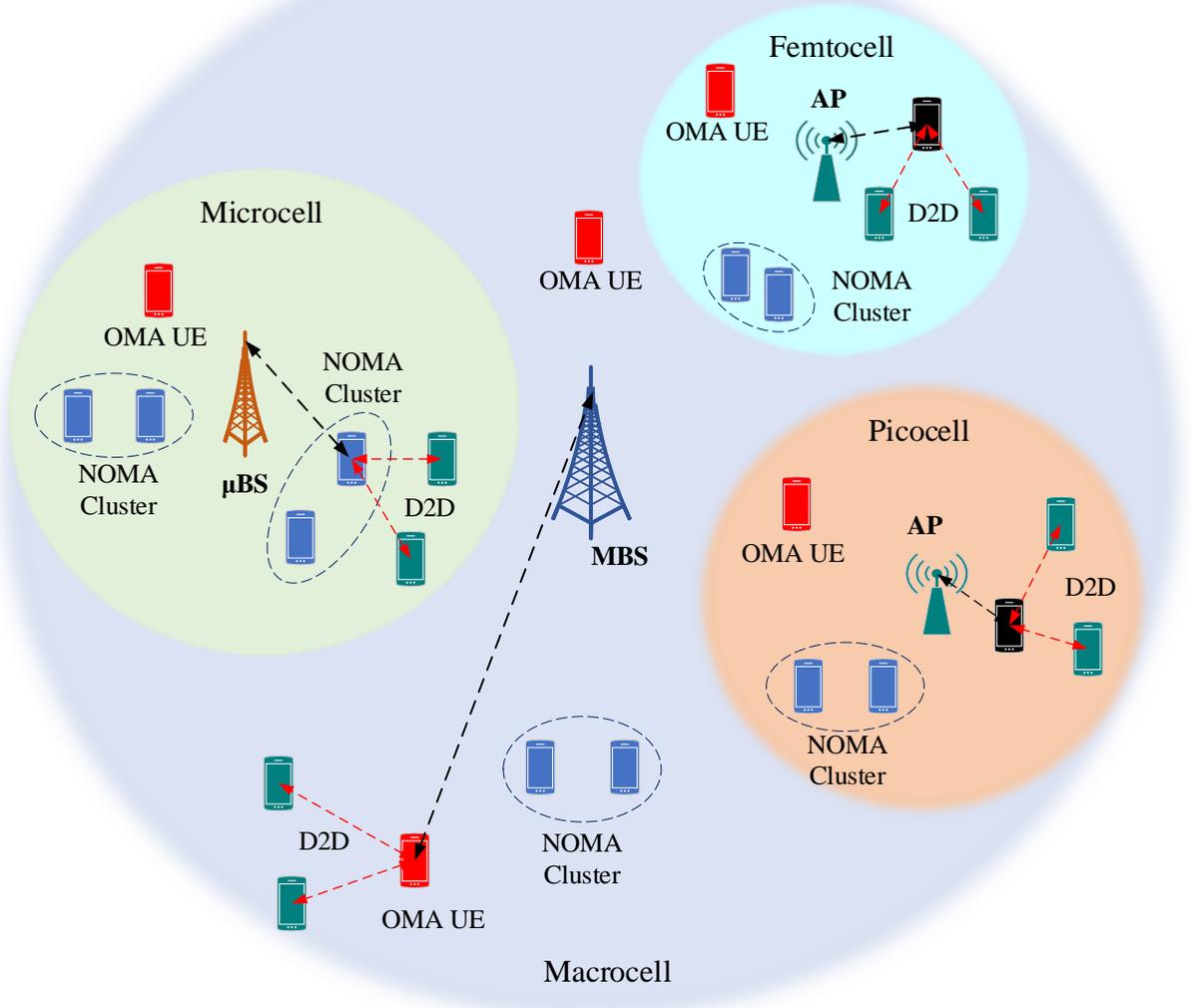

Fig. 1. System model for 6G HetNet.

*B. Multi-layer Architecture for Distributed Learning*

The MUSIC architecture, as shown in Fig. 2, is derived by strategically introducing the computational servers into the HetNet. We propose four layers based on the servers' computational power such that the computational power decreases with each subsequent layer in the following order: cloud (data center or cloud servers), fog (fog servers), edge (AP servers), and device (UEs).

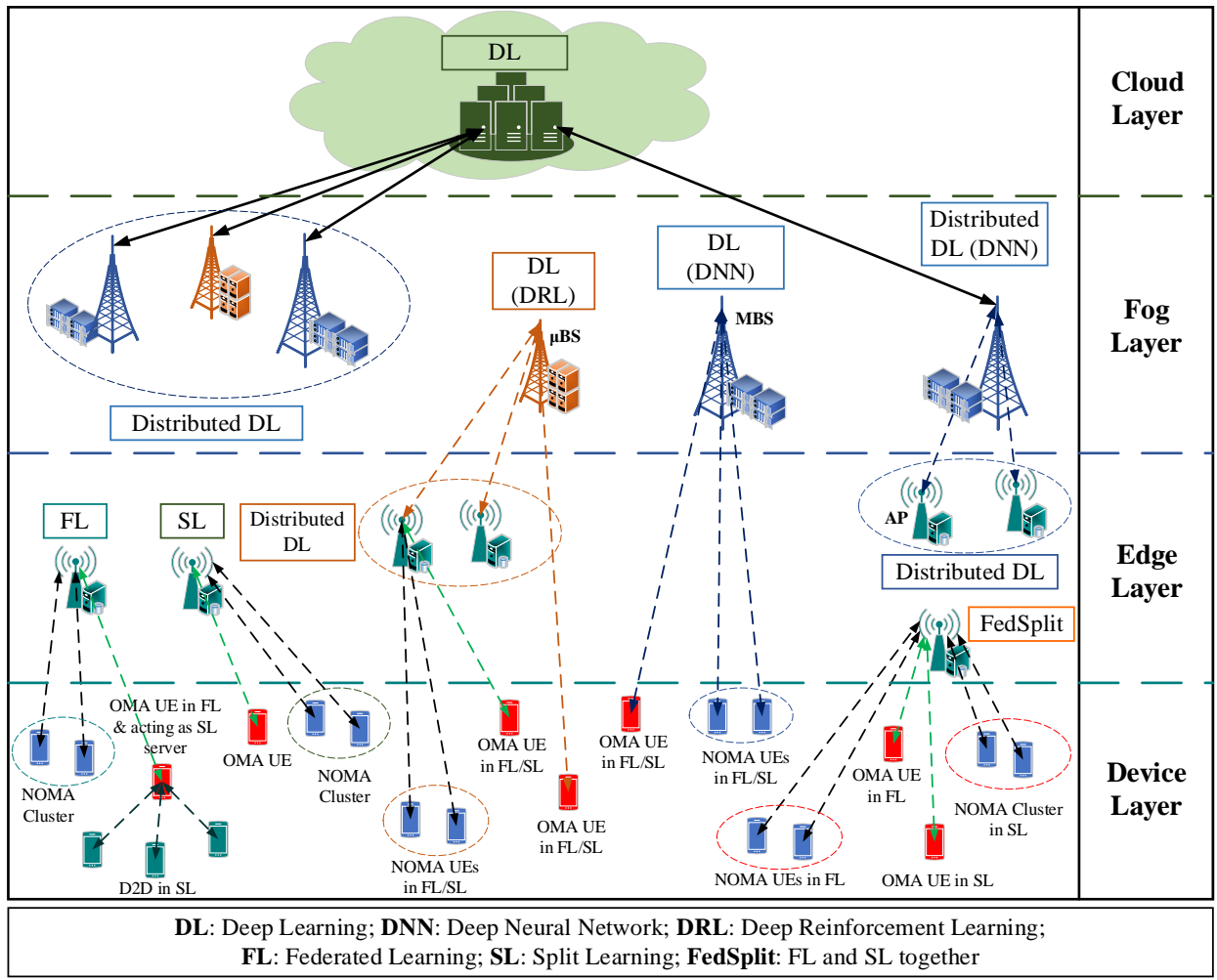

Fig. 2. Proposed multi-layer architecture for distributed ML and MA techniques in 6G.

Thus, at the topmost layer, the cloud layer has the data centers, which house the powerful cloud servers off-site, not necessarily located within the coverage of a particular BS or AP while the device layer constitutes the least computationally powerful UEs which also have the least power flexibility (limited battery power). Evidently, the most power-hungry servers are located at the data centers and the least power-hungry devices would be at the device layer. We limit the architecture to four layers in order to minimize the communication latency (between the endpoint layers). Additionally, the communication between the layers is primarily facilitated by the optical transport network. When a UE in the coverage area of one cell wishes to establish communications with a UE of another cell, the optical network (not shown) connecting the APs and BSs to the MBS serves as the liaison between cells to facilitate the required routing protocols. It is this same optical network which supports the offloading or distribution of ML tasks across APs/BSs throughout the layers.

*C. Collaborative Learning Strategies in 6G HetNet*

According to our proposed multi-layer architecture, each server in the first three layers has the capability of training and testing an ML model on its own (without distributing the work to other servers). However, it can also participate and collaborate with multiple servers at its own or following layer. Thus, the cloud servers can be connected to BSs of the fog layer and can distribute their ML models among the fog servers so that such fog servers would participate together in collaborative learning. The fog servers can further distribute the ML models received from the cloud servers to the edge servers. However, if an ML task is particularly latency-sensitive, the cloud servers may not distribute the tasks over multiple fog servers if utilizing the powerful cloud servers would be the only way to meet such a deadline and overcome multi-layer induced communication latency and signalling. It is worthy to mention that, to train and test a particular ML model, the cloud, fog, and edge servers can use any of the learning algorithms such as supervised, semi-supervised, unsupervised, and reinforcement learning. We also assume that such servers can employ deep learning (DL) for ML training when high accuracy is required (application-dependent).

Each of the fog servers can train ML models of its own and distribute the models to the edge servers. The edge servers can then further distribute the ML models to the next layer if needed. However, we restrict the edge servers to the following decentralized learning techniques: federated learning (FL), split learning (SL), or a combination thereof (FedSplit) to ensure the security of the

UEs participating in the said learning techniques. These techniques are considered to be decentralized since the UEs do not share their data sets with any server or UE device; the data sets are local to the UEs. In order for fog and edge servers to carry out decentralized learning (which requires the use of UEs), they must be aware of the UEs' current power statuses (battery power) and computational capacities because a particular ML approach may severely deplete a UE's power reserves or even violate latency constraints (due to the relatively minimal computing capacity of the UEs). We assume that 6G protocols will enable periodic UE status updates with battery and computational power information. Thus, a BS or AP should only designate UEs with sufficient computing and battery power to participate in such decentralized learning. In general, FL requires much more UE battery and computational power than does SL; the reasons as to why lie in the way FL and SL fundamentally differ from each other which is discussed in the next section.

There are two important points to make at this point. Firstly, there should not be any distribution of ML tasks across more than three layers to minimize latency and communication overhead. Secondly, several factors must be taken into consideration before a server decides to distribute machine learning tasks. For instance, the total power consumption (resulting from both communication and computation) of a single server must be compared with that of multiple collaborating servers which reside on the same or different layers. The computational, propagation, and communication-induced (signalling and coordination among multiple servers) latency of distributing tasks across lower layers must also be considered. A robust decision-making algorithm (not within the scope of this work) should be able to optimize when to distribute tasks such that the latency, security, energy consumption requirements and objectives are met.

## III. FEDERATED AND SPLIT LEARNING TECHNIQUES

In this section, we discuss several distributed collaborative learning approaches involved in the device layer.

### A. Federated Learning Technique

In FL, multiple clients (device layer UEs) collaborate with the FL server in training a complete ML model utilizing their local data sets. Any training done at the client side is referred to as local training (for a local model) while any training done at the server side is known as global training (for a global model). When the UEs complete their local training, they send only the resulting learning parameters to the server; every UE performs multiple local training (local iterations) to obtain its learning parameters which are then sent to the server. This server then aggregates all the learning parameters from all the collaborating clients [12]. After this is done, a single global set of parameters is generated and distributed back to the clients for further local training. Each time the server distributes its global model (aggregated set of the clients' learning parameters) back to the clients, a global iteration has been completed. This process is repeated until the ML model converges and achieves a certain accuracy. Hence, the total duration of FL is the summation of the duration of the global iterations; it is affected by the UE which takes the longest to complete its local model training. Note that no exchange of raw data occurs to preserve the security of the UEs and the computational load is shared among the server and clients; this is the case with any decentralized learning technique.

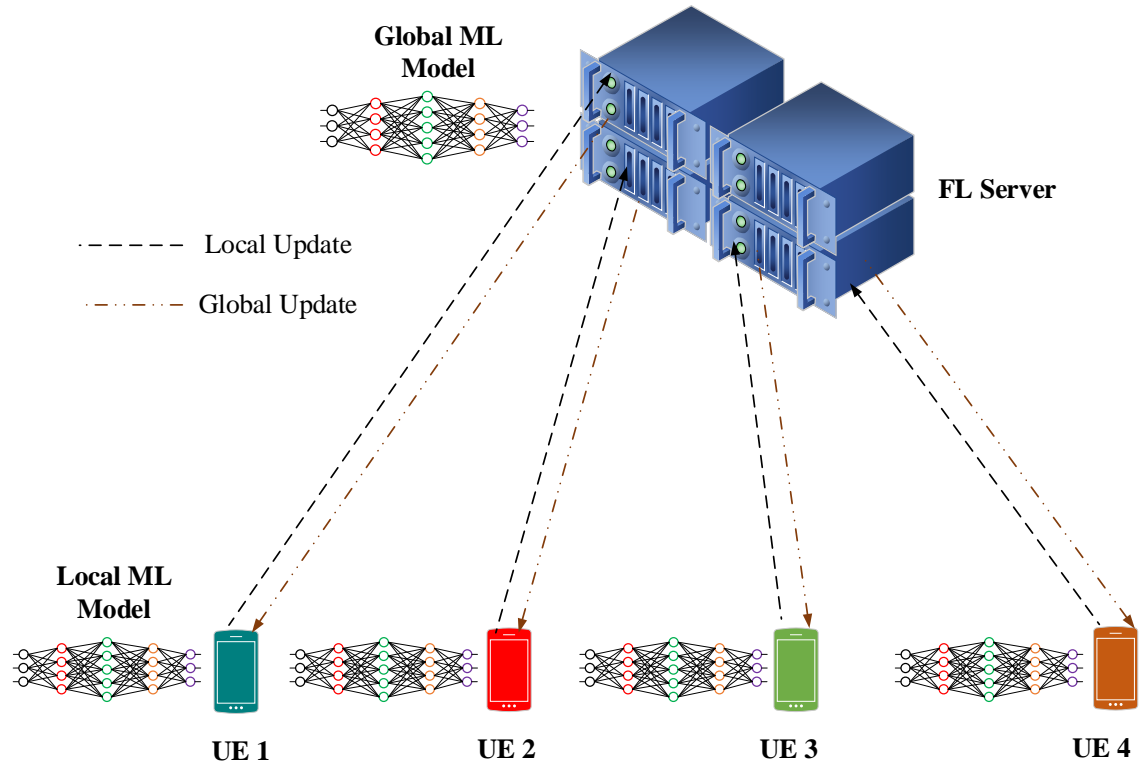

Fig. 3. An overview of FL technique.

## B. Split Learning Technique

We now discuss SL, another distributed collaborative ML technique, which itself has two variations: homogeneous and heterogeneous SL. In both types of SL, each client trains only a portion of the full ML model as opposed to FL where each client trains the full portion of the model. SL, as the name implies, splits the model between the server and clients. The part or layer (for DNN) of the model up to which a client locally trains is called the cut layer. Clients send only the updated weights (also known as smashed data) of their portion of the neural network, specifically, from the cut layer to the SL server [13].

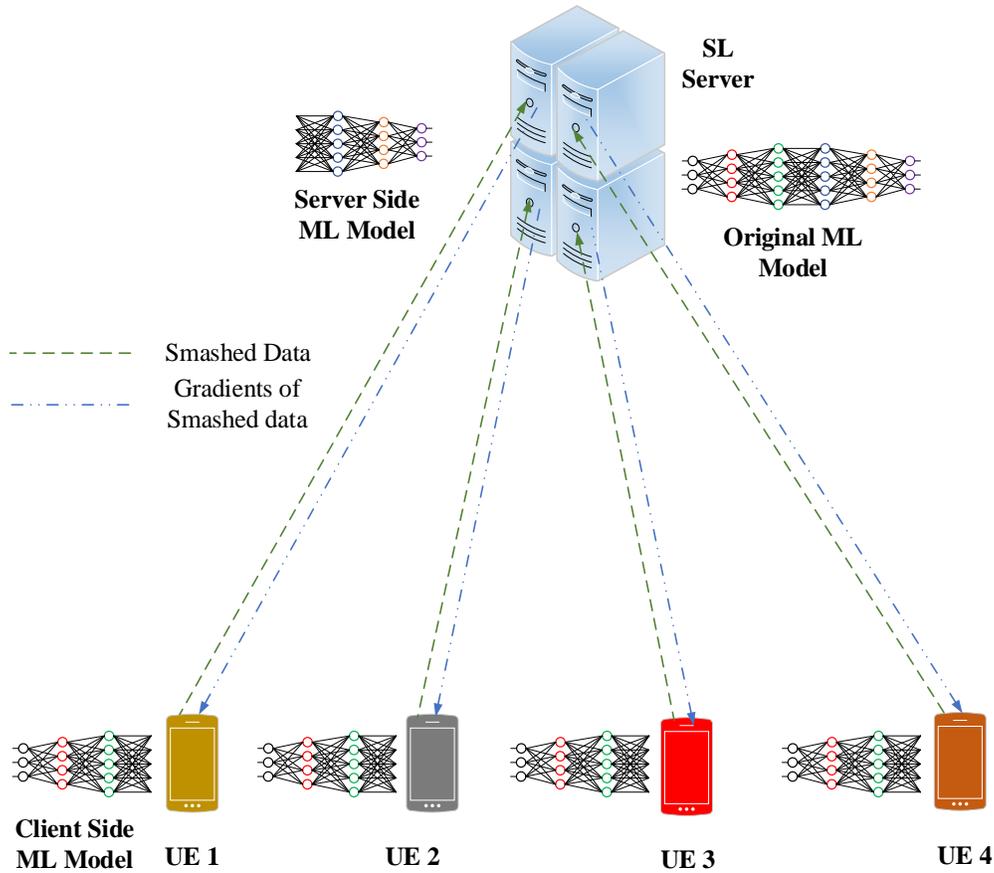

Fig. 4. An illustration of homogeneous SL.

### a) Homogeneous SL Technique

In Fig. 4, we illustrate the homogeneous SL technique where the client portion of the model is the same for all the clients. In this approach, the training takes on a sequential approach as follows. First, a client trains its portion of the model up to the cut layer and then transmits its smashed data to the server. The server then trains its portion of the model with the received smashed data and finishes the forward propagation of the ML network. Afterward, the server performs the back-propagation up to the cut layer and transmits the gradients of the smashed data to the client. The client then executes the back-propagation from the cut layer to the first layer with the received gradients. This completes one single iteration of the training. The SL server then sends the resultant learning parameters to the next client which repeats the above process to complete another iteration. Evidently, each client except the one currently locally training will be idle until its own turn has arrived. This process continues until the ML model converges.

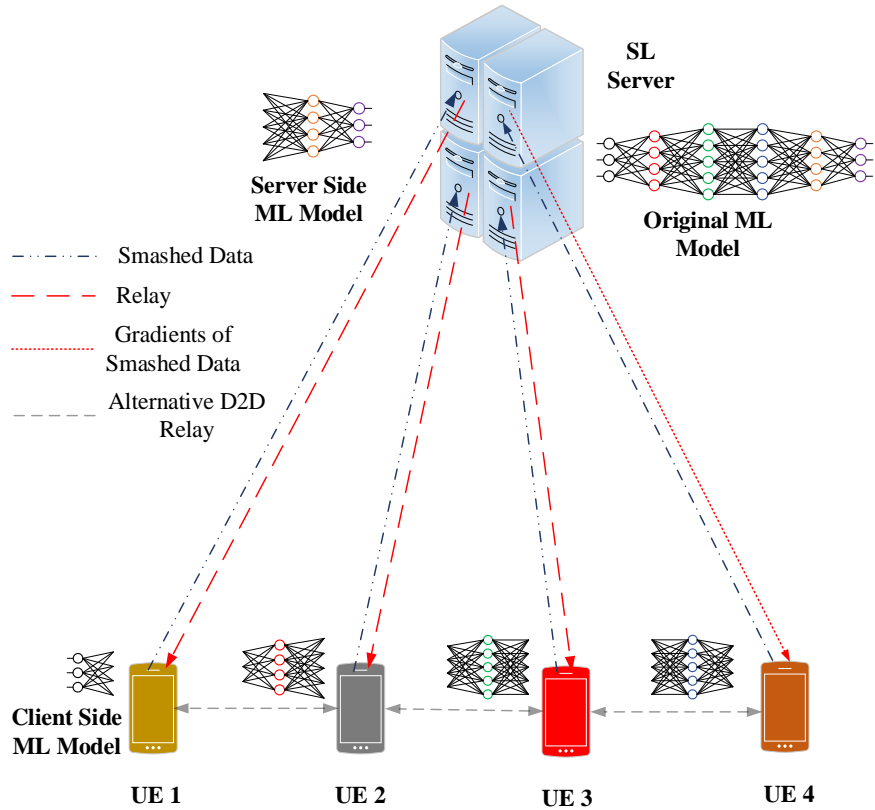

Fig. 5. An overview of heterogeneous SL.

In Fig. 5, which demonstrates the heterogeneous SL technique, the SL server again splits the entire model between the client side and server side as in homogeneous SL. However, at the client side, each client trains a unique portion of the overall client-side model, i.e, the cut layer for each client is different. All the clients perform their local training according to their cut layer sequence; the client with the last cut layer sends its final smashed data to the server. Then, the server will train its side of the model and this completes the forward propagation of the model. The server then performs back-propagation on its model and sends gradients of the smashed data to the client with the last cut layer. Now, each of the clients performs back-propagation on its own layer(s) in the reverse sequence of the forward propagation and completes the back-propagation of the model training. The whole process completes a single iteration of this SL learning variation. We should state here that each client conventionally would forward or relay its smashed data to the subsequent UE by way of the SL server. In cases where D2D networks are involved, the UE can omit the SL server as an intermediary node and simply relay or forward its smashed data directly to the next UE. Thus, relay simply means forwarding the smashed data in the context of the figure. Note that only one client participates in a single iteration in homogeneous SL while in heterogeneous SL, every client participates in a single iteration. In other words, a single iteration is completed each time there is a pair of forward and backward propagation over the entire model; this obviously is only possible when all the clients are involved. The training duration of heterogeneous SL is much longer as compared to that of homogeneous SL since each UE trains only a small and unique portion of the client-side model and does so sequentially. However, the training latency can be shortened further if each client in heterogeneous SL can participate in a D2D network and send updates directly to the next client rather than going through the server as a relay. In this case, the clients have to be situated within close proximity of each other to form a D2D network.

It is worth noting here that as shown in the device layer of MUSIC in Fig. 2, SL can occur as a sub-learning process in an FL session as depicted on the extreme left of the device layer. For example, there can be multiple UEs participating in an FL session, and one such UE can distribute its own local FL training over multiple D2D UEs by way of SL. In this case, the master UE (intermediate UE) acts as an SL server. From the perspective of the edge server, it is only one UE that sent its update (not all the UEs that participated in the D2D SL). In other words, the slave UEs in the D2D network are invisible to the fog server. Such a distribution of tasks may be employed if latency requirements are not stringent and if a particular UE does not possess the computational capacity or battery power to participate in an FL session with the edge server; thus, a UE can distribute such tasks via SL to the D2D devices underneath. For D2D mesh networks which participate in ML training, there are several fully decentralized distributed ML frameworks that mitigate issues such as redundant communications, updates and the lack of synchronization between UEs. Such frameworks do not necessarily have to utilize the conventional master-slave relationship.

There arise complications however when utilizing SL for mobile UEs; poor wireless channels may impact the overall accuracy of SL. Specifically, as SL is very dependent on a specific sequence of UEs participated in training the ML model, if a UE is dropped from the training iteration due to poor connectivity, it will hamper the entire SL process. Thus, the UE selection for SL should be meticulous enough such that the probability of stragglers will be minimized. Only UEs which are either experiencing relatively invariant channel conditions or are immobile (such as when charging) should be allowed to participate; this ensures that the advantages of SL over FL with respect to energy efficiency can indeed be exploited.

IV.    MULTIPLE ACCESS SCHEMES AND D2D FOR COLLABORATIVE LEARNING

In Fig. 6, we detail the multiple access schemes from a purely communication-based perspective within a single cell at the device layer. The two well-known MA schemes are NOMA and OMA; both can utilize what can be referred to as grant-based and grant-free schemes. Grant-free schemes relieve the need of signalling between a UE and BS for UL transmission, thus reducing the communication latency. In this fashion, decentralized learning techniques can be shortened in duration via grant-free schemes. When each client in decentralized techniques needs to send its updates to the servers, grant-free schemes will enable the client to send its updates on demand as needed.

NOMA allows a set of UEs, referred to as NOMA clusters, to be assigned the same frequency resources by the AP. Although the inner workings of cluster formation are out of the scope of this work, it should be noted that UEs need not be within close proximity of each other to be assigned to the same cluster. In this fashion, NOMA allows more users to participate in collaborative learning since they can share resources and minimize queuing delays (overall reduction in communication delays).

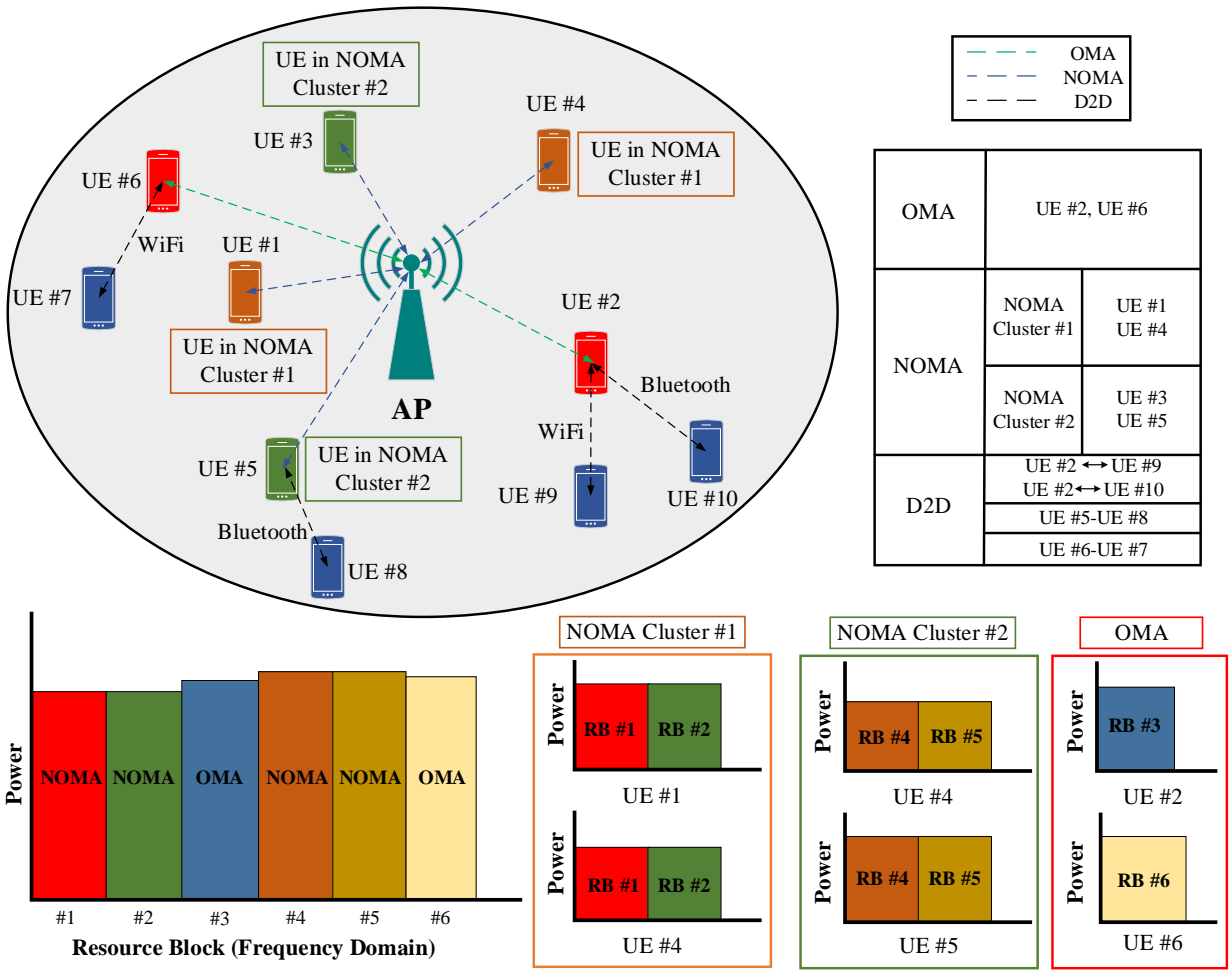

Fig. 6. Multiple access schemes for communications in distributed collaborative learning.

In Fig. 6, special attention should be paid to the frequency and power allocation to the participating UEs. Note that in the first NOMA cluster, there are two UEs sharing the same pair of resource blocks, but each is transmitting with a different power so that the receiver can distinguish between the two UE signals. The same is also true for the pair of UEs in the second NOMA cluster; they too share the same resource blocks but with different power. In general, we can visually observe that NOMA is spectrally efficient since it allows the simultaneous usage of frequency resources by multiple UEs as opposed to OMA where singular access is key. With respect to D2D communications, i.e., Bluetooth or WiFi link, only the UE which serves as an intermediary between the D2D devices and the AP will be allocated resources by the AP via either NOMA or OMA. Furthermore, we stipulate that D2D networks should be only single hop after the intermediate UE since each intermediate UE distributes learning tasks to the UEs below it. Having multiple UEs distributing tasks in each subsequent hop complicates such learning tasks and master-slave relationship.

## V. RESEARCH ISSUES

The extremely heterogeneous and highly stringent QoS requirements of 6G create several issues and challenges to tackle in the near future. End-to-end cross domain resource orchestration is a great challenge particularly with respect to the "curse of dimensionality" problem which is daunting even for the most computationally heavy-duty ML servers. Moreover, much of the decision making must be done in real-time. ML algorithms are initially trained offline to accelerate the online decision making, but it often leads to policies and decisions that are overfitted for the offline environments. Lastly, adversarial ML always poses a challenge that can wreak havoc on network management. However, the major drawbacks and significant research issues related to our proposed MUSIC are pointed out below.

## A. Optimal UE Pool Selection

Observably, many of the ML techniques rely very heavily in one form or another, on an optimal UE pool selection. UEs are very heterogeneous in their computational capabilities and a single UE can bottleneck an entire collaborative training process (such as that of FL, SL, etc.), especially if such a UE has very poor channel conditions, computational power, battery power, etc. There are several implied restrictions as to which UE should ideally be able to participate in collaborative learning due to the above-mentioned constraints. Moreover, a malicious UE may send harmful ML parameters to the FL or SL servers that will impact, via a cascaded effect, resource orchestrations being executed at other domains or ends of the network. This is one of the major drawbacks with such an inter-connected ML architecture such as that of MUSIC.

## B. Dynamic Architecture vs. ML Accuracy

The highly dynamic architecture of 6G networks poses a vital challenge for deploying ML in general. The dynamic nature of 6G network configurations needs to be taken into serious consideration before deploying ML techniques to intelligentize the networks. The constant reconfiguration of future generation networks may deteriorate the accuracy of the ML models [7]. For example, variations in QoS requirements of heterogeneous applications will produce data sets with enormous variations; such data sets are not adequate for training different ML models to predict network conditions and enhance network functions (channel estimation, resource allocation, channel coding and decoding, etc.). Not only that but obtaining higher accuracy will also undoubtedly be an issue.

*C. Overhead vs. AI Performance*

The distributed multi-layer architecture and various ML techniques also introduce significant computing and communication overheads in 6G networks. The management of the network entities involved in distributed ML along with different MA schemes requires a holistic overview to obtain satisfactory performance in an AI-enabled 6G HetNet architecture. Moreover, the communication signalling within 6G networks must be done parallelly with ML-related transmissions. This is on top of the traditional communication load generated by the traditional mobile network users who do not necessarily participate in ML training and testing [14].

*D. Network and ML Security*

The distributed architecture of ML exposes a big concern for 6G security, specifically that of the ML participants in distributed learning [5]. There needs to be a set of feasible solutions which can address the security issues related to the distribution of ML models as well as adversarial ML which seeks to cause malicious ML training leading to poor network performance, failures, and security lapses in network servers and UEs.

*E. Optimal Decision-Making for Collaborative ML*

There needs to be a rigorous framework which dictates or assists the 6G network in deciding when to distribute ML tasks over multiple servers, layers, and UEs. A management entity comparable to that of a software-defined-networking controller should be aware of the latency trade-offs, computational and power costs of any course of action (distribution of ML tasks). The overall performance, spectral and energy efficiencies of a layer-based architecture are highly dependent on such decisions [2].

*F.  Power and Spectral Resource Allocation*

The heterogeneity of 6G applications poses a significant challenge for BSs and APs in allocating power and spectral resources to the UEs. The utilization of several transmission bands simultaneously, extreme mobility of UEs, and channel conditions at such high frequencies (mmWave and THz bands) create spectral efficiency issues [9]. They also impact the transmission power allocation of the UEs at the access end of the wireless networks. Cumulatively, all the above mentioned challenges affect the overall performance of AI at the device layer (the energy consumption and training time of the clients may increase) as well as the upper layers in the proposed 6G HetNet. To address such issues, efficient resource allocation is vital; however, doing so on an individualistic level for each user does not guarantee an efficient allocation for the entire system. Integrating game theoretic approaches with machine learning may provide better solutions [15].

*G.  Learning Models*

Another important issue for the future generation 6G networks is the utilization of effective learning models for the prediction of network functions and conditions to optimally provision for the extremely heterogeneous envisioned applications. This is so because not every ML model is suitable for every scenario especially under rapidly evolving network conditions. The distribution of the proper ML model is crucial due to the need of high accuracy when the intelligent 6G network needs to make decisions regarding the network configuration, interpretation of unlabeled data sets, and so on [5].

*H.  Computational Complexity of ML Models*

The relation between the computational complexity and ML training requires optimal management of the computing resources of the servers and UEs. A highly complex model can reduce the computational power efficiency and further lead to deadline violations. For instance, splitting a model in SL between the clients and a server introduces a trade-off between training time and computational energy consumption. Furthermore, the accuracy of such a learning technique will be greatly affected by the partitioning of the model between the server and the client (i.e., how many layers are to be trained by the server vs. how many are to be trained by the clients). The training time is further elongated when more UEs are involved in SL (although this increases accuracy). All the above must be considered when partitioning the computational resources of a server and UEs for particular ML tasks.

*I.  Communication Overhead*

The communication overhead induced by the distribution of an ML task across multiple layers, frequent UE transmissions of ML parameters to the servers (and vice versa) during the training sessions of collaborative learning, the repeated transmissions of such (in case of failed transmissions), and the respective acknowledgments are additional factors to be considered in a distributed machine learning architecture. Resultantly, the network latency tends to suffer. Special attention should be paid to exploring avenues to reduce the communication overhead incurred by the aforementioned issues. Additionally, sophisticated interference reduction mechanisms need to be adopted to mitigate the hurdles of wireless bandwidth sharing by multiple UEs.

# VI.  CONCLUSION

The layer-based distributed ML approaches will be quite central in intelligentizing the future generation 6G HetNet; they will allow the network to execute different network functionalities and adapt to dynamic scenarios efficiently while reducing the overall energy consumption as well as enhancing the spectral and computational efficiencies of said ML techniques. While it may be true that the overhead of communications and the associated costs will increase due to the multilayer nature of the ML hierarchy presented herein, the communication costs are significantly lower than that of the computational tasks. In addition, the integration of different MA schemes with collaborative learning serves to further minimize the energy costs and computational loads on not only the 6G HetNet but the UEs as well. In this work, we have illustrated different types of decentralized ML approaches which are distributed over several network entities and layers of our proposed MUSIC. We have also discussed the advantages of utilizing D2D communications and the grant-free scheme in order to shorten the latency of such ML training sessions.

**Mohammad Arif Hossain** [S'19] is currently pursuing his Ph.D. degree in Computer Engineering at the New Jersey Institute of Technology, USA. He has (co-)authored more than thirty articles and contributed to some IEEE standardization groups. His current research interests include distributed machine learning, cloud computing, and future generation wireless networks.

**Abdullah Ridwan Hossain** [S'20] is currently pursuing his doctorate in Electrical Engineering at the New Jersey Institute of Technology. He has (co-)authored ten publications as well as three book chapters. His research interests include machine learning, optimization of wireless and aerial communications, as well as accreditation and assessment.

**Nirwan Ansari** [S'78, M'83, SM'94, F'09], Distinguished Professor of Electrical and Computer Engineering at the New Jersey Institute of Technology, has (co-)authored three books and many articles. He has guest-edited a number of special issues covering various emerging topics in communications and networking. He has served on the editorial/advisory board of over ten journals including as Associate Editor-in-Chief of IEEE Wireless Communications Magazine. His current research focuses on green communications and networking, cloud computing, drone-assisted networking, and various aspects of broadband networks.